\documentstyle[12pt]{article}

\def\a{\alpha }

\begin{document}

\begin{center}
{\Large\bf NLO QCD Analysis of Polarized Deep Inelastic Scattering}
\end{center}
\vskip 2cm

\begin{center}
{\bf Elliot Leader}\\
{\it Department of Physics\\
Birkbeck College, University of London\\
Malet Street, London WC1E 7HX, England\\
E-mail: e.leader@physics.bbk.ac.uk}\\
\vskip 0.5cm
{\bf Aleksander V. Sidorov}\\
{\it Bogoliubov Theoretical Laboratory\\
 Joint Institute for Nuclear Research\\
141980 Dubna, Russia\\
 E-mail: sidorov@thsun1.jinr.dubna.su}
\vskip 0.5cm
{\bf Dimiter B. Stamenov \\
{\it Institute for Nuclear Research and Nuclear Energy\\
Bulgarian Academy of Sciences\\
Blvd. Tsarigradsko chaussee 72, Sofia 1784, Bulgaria\\
E-mail:stamenov@inrne.acad.bg }}
\end{center}

\vskip 0.3cm
\begin{abstract}
We have carried out a NLO analysis of the world data on polarized
DIS in the $\overline{MS}$ scheme. We have studied two models of
the parametrizations of the input parton densities, the first due
to Brodsky, Burkhardt and Schmidt (BBS) which gives a simultaneous
parametrization for both the polarized and unpolarized densities
and in which the counting rules are strictly imposed, the second 
in which the input polarized densities are written in terms of
the unpolarized ones in the generic form $~\Delta q(x)=f(x)q(x)~$ 
with $f(x)$ some simple smooth function. In both cases a good fit
to the polarized data is achieved. As expected the polarized data
do not allow a precise determination of the polarized gluon density.
Concerning the polarized sea-quark densities, these are fairly
well determined in the BBS model because of the interplay of
polarized and unpolarized data, whereas in the second model,
where only the polarized data are relevant, the polarized
sea-quark densities are largely undetermined.
\end{abstract}
\vskip 0.5 cm
\newpage

{\bf 1. Introduction.}
\vskip 4mm

Experiments on polarized deep inelastic scattering (DIS) were
initiated at SLAC by the SLAC-Yale group \cite{Yale} soon after
the discovery of Bjorken scaling. Enormous impetus was given to
the subject by the European Muon Collaboration (EMC) experiment
at CERN \cite{EMC} in 1988 whose results seemed to imply a "spin
crisis in the parton model" \cite {ELA}. Much theoretical and
experimental work has followed and today there is a rich program
of experiments under way (E154, E155 at SLAC; HERMES at HERA)or
in the progress of being set up (COMPASS at CERN).\\

Experiments on unpolarized DIS provide information on the
unpolarized quark densities $q(x,Q^2)$ and  gluon density $G(x,Q^2)$
inside a nucleon. Polarized DIS experiments, using a
longitudinally polarized target, give us more detailed information,
namely the number densities of quarks $~q(x,Q^2)_{\pm}~$ and gluons $~G(x,
Q^2)_{\pm}~$ whose helicity is respectively along or opposite to the 
helicity of the parent nucleon. The usual densities are
\begin{equation}
q(x,Q^2) = q_{+}(x,Q^2) + q_{-}(x,Q^2)~,~~~~  
G(x,Q^2) = G_{+}(x,Q^2) + G_{-}(x,Q^2) 
\label{unpolpa}
\end{equation}
and the new information is then contained in the polarized
structure function $~g_1(x,Q^2)~$ which is expressed in terms of
the {\it polarized} parton densities 
\begin{equation}
\Delta q(x,Q^2) = q_{+}(x,Q^2) - q_{-}(x,Q^2)~,~~~~
\Delta G(x,Q^2) = G_{+}(x,Q^2) - G_{-}(x,Q^2)~. 
\label{polpa}
\end{equation}

Two developments in the past few years have made it possible and
worthwhile to attempt a detailed comparative study of the polarized 
parton distributions. On the one hand, a wealth of new data, much of it
high quality, has appeared [4 - 12]. On the other, the
theoretical calculation of the Altarelli-Parisi splitting
functions to two-loop order has been, after several hiccoughs,
completed successfully \cite{nlocor}.\\

It would be wrong, however, to imagine that the polarized
densities can now be determined to the same accuracy with which
the unpolarized densities are known. This can be understood quite
simply. Up to the present the polarized data consist solely of fully
inclusive neutral current (in effect, photon induced) reactions
on protons and (via deuterium or Helium-3) on neutrons, {\it
i.e.} one has information on the two polarized structure
functions $~g_1^{p}(x,Q^2)~$ and $~g_1^{n}(x,Q^2)~$. Even if one
makes some simplifying assumptions about the polarized sea, e.g.
$~\Delta\bar{u}(x) = \Delta\bar{d}(x) = \Delta\bar{s}(x)~$ or 
$~\Delta\bar{u}(x) = \Delta\bar{d}(x) = 2\Delta\bar{s}(x)~$ one is still
expressing two experimental functions in terms of four densities:
$~\Delta u(x,Q^2),~ \Delta d(x,Q^2),~ \Delta\bar{q}(x,Q^2)~$ and 
$~\Delta G(x,Q^2)$.
What is lacking here is the information from charged current
reactions which plays an important role in pinning down the
unpolarized densities. The situation is alleviated by the
beautiful connection between the first moments of the polarized
parton densities and weak interaction physics. Namely, one has
the connection with neutron $\beta~$-decay, via the Bjorken sum rule,
\begin{equation}
\int_{0}^{1}dx[\Delta u(x,Q^2)+\Delta\bar{u}(x,Q^2)-
\Delta d(x,Q^2)-\Delta\bar{d}(x,Q^2)]=g_{A}/g_{V}  
\label{BSR}
\end{equation}
and, to the extent that flavour $~SU(3)~$ is a good symmetry, the
connection with hyperon $~\beta~$-decay
\begin{eqnarray}
\nonumber
\int_{0}^{1}dx\{\Delta u(x,Q^2)+\Delta\bar{u}(x,Q^2)
&+&\Delta d(x,Q^2)+\Delta\bar{d}(x,Q^2)\\
&-&2[\Delta s(x,Q^2)+\Delta\bar{s}(x,Q^2)]\}=3F-D.  
\label{BHSR}
\end{eqnarray}

The values of $~g_{A}/g_{V}~$ and 3F-D are taken from \cite{PDG}
\begin{equation}
g_{A}/g_{V}=1.2573~\pm~0.0028,~~~~3F-D=0.579~\pm~0.025~.
\label{GA3FD}
\end{equation}

Assuming a roughly flavour-independent polarized sea, allows one
to interpret Eqs. (\ref{BSR}) and (\ref{BHSR}) as statements about 
the first moments of the polarized valence densities: 
\begin{equation}
\int_{0}^{1}dx[\Delta u_{v}(x,Q^2)-\Delta d_{v}(x,Q^2)]\cong 1.26~,  
\label{BSRa}
\end{equation}
\begin{equation}
\int_{0}^{1}dx[\Delta u_{v}(x,Q^2)+ \Delta d_{v}(x,Q^2)]\cong 0.58~,
\label{BHSRa}
\end{equation}
which immediately  suggests that $~\Delta d_{v}(x,Q^2)~$ is of
opposite sign to $~\Delta u_{v}(x,Q^2)~$ and of roughly comparable
magnitude, in agreement with simple $~SU(6)~$ models of the proton
wave-function. Eqs. (\ref{BSR}) and (\ref{BHSR}) are crucial
supplements to the polarized DIS data.\\

In seeking input parametrizations of the polarized densities
into the QCD evolution equations one must clearly respect the
positivity of the number densities $~q(x,Q^2_{0})_{\pm}~$,
which {\it via} (\ref {unpolpa}) and (\ref {polpa}) is equivalent to 
demanding
\begin{equation}
\vert {\Delta q(x,Q^2_{0})} \vert\leq q(x,Q^2_{0})
\label{posit}
\end{equation} 
  
There are, in addition, certain {\it counting rules} relating to
the behaviour of $~\Delta q(x)/ q(x)~$ as $~x\to 0~$ and 
$~x\to 1~$, which follow in the parton model from perturbative
QCD and the form of the infinite momentum frame nucleon wave function
[15 - 17].\\ 

We have examined two classes of models for the input densities. The
first, due to Brodsky, Burkardt and Schmidt \cite {BBS} is
unusual since it directly parameterizes $~q_{+}(x,Q^2)~$ and
$~q_{-}(x,Q^2)~$ at some $~Q^2=Q^2_0~$ (rather than $~\Delta 
q(x,Q^2_0)~$
and $~q(x,Q^2_0)~$) so that the free parameters are determined
from a simultaneous fit to the polarized and unpolarized data.
Positivity is simple to implement and the counting rules are
imposed exactly at $~Q^2=Q^2_0~$. 

In the second, each polarized parton distribution is written in
the generic form $~\Delta q(x,Q^2_0)=f(x)q(x,Q^2_0)~$ at
some $~Q^2_0~$, with the usual densities $~q(x,Q^2_0)~$
determined from the unpolarized data (in practice we utilize the
MRS(A\'{}) set of distributions \cite{MRS}). The functions $~f(x)~$
are parameterized so as to respect positivity but the counting
rules are not imposed in a strict fashion. 

Finally it should be remembered that beyond the leading order in
perturbation theory the parton densities become scheme dependent.
In this paper we work in the $\overline{MS}$ scheme.\\

In Section 2 we wish to draw the reader's attention to certain
interesting qualitative features of the polarized DIS data. In
Section 3 we explain the method of analysis 
and in Section 4 discuss the parametrization of the models and
their properties. Our results are presented and analyzed in
Section 5 and conclusions follow in Section 6.\\

{\bf 2. Implications of qualitative features of the data}
\vskip 4mm

The structure functions $~g_1^{p,n}(x,Q^2)~$ are determined in the
following $x$ range:$~0.003 < x < 0.8~$. For small $x$ there
are large errors on the measured values, but if one takes the
central values of the data points as indication of the trend of
the behaviour as $~x\to 0~$ then one is led to some surprising
conclusions. To see this, consider the expressions for the
structure functions $~g_1^{p,n}(x,Q^2)~$. They can be expressed
in terms of contributions $~\Delta q_3(x,Q^2),~\Delta q_8(x,Q^2)~$ 
and $~\Delta \Sigma(x,Q^2)~$ of definite flavour symmetry as 
\begin{equation}
g_1^{p,n}(x,Q^2)=\{\pm {1\over 12}\Delta q_3(x,Q^2) + 
{1\over 36}\Delta q_8(x,Q^2) + 
{1\over 9}\Delta \Sigma(x,Q^2)\}\{1 + O(\alpha_s)\}
\label{g1pn}
\end{equation} 
where the flavour non-singlet contributions are
\begin{equation}
\Delta q_3(x,Q^2)=\Delta u(x,Q^2)+\Delta\bar{u}(x,Q^2)-
\Delta d(x,Q^2)-\Delta\bar{d}(x,Q^2)~,
\label{a3x}
\end{equation} 
\begin{eqnarray}
\nonumber
\Delta q_8(x,Q^2)=\Delta u(x,Q^2)+\Delta\bar{u}(x,Q^2)
&+&\Delta d(x,Q^2)+\Delta\bar{d}(x,Q^2)\\
&-&2[\Delta s(x,Q^2)+\Delta\bar{s}(x,Q^2)]
\label{a8x}
\end{eqnarray} 
and the singlet contribution is
\begin{equation}
\Delta \Sigma(x,Q^2)=\sum _{f}[\Delta q_f(x,Q^2)+\Delta 
\bar{q}_f(x,Q^2)]~.
\label{dsig}
\end{equation} 

The gluon contribution to $~g_1(x,Q^2)~$ is hidden in the
$~O(\alpha_s)~$ correction terms in (\ref{g1pn}) and a more precise
expression will be given in Eq. (\ref{NLOmomg1}).

One sees then that the difference $~g_1^{p}(x,Q^2)-g_1^{n}
(x,Q^2)~$ is a purely non-singlet, whereas the sum $~g_1^{p}(x,Q^2)
+g_1^{n}(x,Q^2)~$ is a mixture of singlet and non-singlet
contributions.

Now according either to the small-$x$ behaviour of the evolution
equations \cite{QCDsmallx} or to the summation of double logarithmic
terms at small $x$ \cite{Ryskin}, the singlet contribution should
dominate over the non-singlet terms as $~x\to 0~$, which, by the
above, could imply ($~g_1^{p}+g_1^{n})~>~(g_1^{p}-g_1^{n})~$ at
small $x$. The data (see Fig. 1) show precisely the opposite trend.
However, while the theoretical arguments predict the form of the
behaviour as $~x\to 0~$, namely $~C_{ns}x^{-a_{ns}}~$ or
$~C_{s}x^{-a_{s}}~$ with $~a_s>a_{ns}>~0$, the values of the
coefficients $~C_{ns},~C_s~$ are sensitive to the structure of
the parton distributions at $~Q^2_0$. So there need not be a
contradiction at presently measured $x$-values. However, as
experiments probe smaller and smaller $x$, there ought to be a
dramatic change in the trend of the data or else the theoretical
arguments are incomplete.

On a practical level concerning the present data, the behaviour
in Fig. 1 can lead to difficulties when one makes, as one is
forced to do, some simplifying assumption for the polarized sea,
such as $~\Delta \bar{u}(x,Q^2_0)=\Delta\bar{d}(x,Q^2_0)$.
For then $~\Delta q_3(x,Q^2)~$ in (\ref{a3x}) practically depends only
on valence distributions and in some cases one finds that the 
best-fit parameters tend to make $~\Delta d_v(x,Q^2)~$ so large
and negative at small $x$ that positivity is violated.\\   

{\bf 3. Method of analysis}
\vskip 4mm
Measurements of polarized deep inelastic lepton nucleon
scattering yield {\it direct} information on the virtual photon-
nucleon asymmetry $~A_1^{N}(x,Q^2)~$. Neglecting as usual the
subdominant contributions, $~A_1^{N}(x,Q^2)~$ can be expressed
{\it via} the polarized structure function $~g_1^{N}(x,Q^2)~$ as
\begin{equation}
A_1^{N}(x,Q^2)\cong {g_1^{N}(x,Q^2)\over F_1^{N}(x,Q^2)}=
{g_1^{N}(x,Q^2)\over F_2^{N}(x,Q^2)}[2x(1+R^{N}(x,Q^2)]~, 
\label{assym}
\end{equation} 
where
\begin{equation}
R^N = (F_2^N-2xF^N_1)/2xF^N_1
\label{ratio}
\end{equation} 
and $~F_1^N~$ and $~F_2^N~$ are the unpolarized structure functions.

Usually the theoretical analysis of the data is presented in
terms of $~g_1^{N}(x,Q^2)~$ extracted from the measured values of
$~A_1^{N}(x,Q^2)~$ according to (\ref{assym}) using different
parametrizations of the experimental data for $F_2$ and $R$. The
parametrizations \cite{NMC} of $F_2$ and \cite{SLAC} of $R$  were
used in the most recent analyses. Some experimental groups assume
also $~Q^2~$ scaling of $~A_1^{N}(x,Q^2)~$ in their extraction of 
$~g_1^{N}(x,Q^2)~$. However, bearing in mind the recent NLO
calculations of $~g_1~$ in QCD, this assumption is not
theoretically correct, especially in the small $x$ region.\\

In our analysis we follow the approach first used in \cite{Vog},
in which the next-to-leading (NLO) QCD predictions
for the spin-asymmetry $~A_1^{N}(x,Q^2)~$ are confronted with the
{\it directly measured} values of $~A_1^{N}(x,Q^2)~$ rather than
with the $~g_1^{N}(x,Q^2)~$ derived  by the procedure mentioned above.
A further advantage of such an approach is that higher twist
contributions are expected to partly cancel in the ratio (\ref{assym}),
in contrast to the situation for $~g_1^{N}(x,Q^2)~$. Usually to
avoid the influence of higher twist effects $~Q^2$-cuts in the 
$~(x,Q^2)~$ data set are introduced. Bearing in mind that in 
polarized DIS most of the small $x$ data points are at low $~Q^2~$, a 
lower than usual 
cut is needed ($~Q^2>1~GeV^2~$) in order to have enough
data for a theoretical analysis. We consider that in this 
approach such a low $~Q^2$-cut is more reasonable.

In NLO approximation 
\begin{equation}
A_1^{N}(x,Q^2)_{NLO}\cong {2xg_1^{N}(x,Q^2)_{NLO}\over 
2xF_1^{N}(x,Q^2)_{NLO}}.
\label{NLOas}
\end{equation} 
In (\ref{NLOas}) $~N=p,~n~$ and $~d=(p+n)/2~$.

To calculate $~g_1^{N}(x,Q^2)~$ and $~2xF_1^{N}(x,Q^2)~$ we have
used the analytic NLO solutions for the moments in Mellin 
space with the $n$th moment being defined by 
\begin{equation}
M_n^{N}(Q^2)=\int_{0}^{1}dxx^{n-2}xg_{1}^N(x,Q^2)~,~~~~~n=1,2,...
\label{momg1}
\end{equation}
\begin{equation}
\overline{M}_n^{N}(Q^2)=\int_{0}^{1}dxx^{n-2}2xF_{1}(x,Q^2)~,~~~~n=2,3,...
\label{momF1}
\end{equation}

In NLO approximation with $~n_f=3~$ active flavours the moments (\ref{momg1})
of the structure function $~g_1^{N}(x,Q^2)~$ can be written in
the form \cite{Stam}:
\begin{eqnarray}
\nonumber
M_n^{p(n)}(Q^2)&=&\{\pm {1\over 12}\Delta q_3(n,Q^2) + 
{1\over 36}\Delta q_8(n,Q^2) + 
{1\over 9} \Delta \Sigma(n,Q^2)\}_{NLO}\\
\nonumber
&+&{\alpha(Q^2)\over 2\pi}\delta C^q_n
\{\pm {1\over 12}\Delta q_3(n,Q^2) + {1\over 36}\Delta q_8(n,Q^2) 
+ {1\over 9} \Delta \Sigma(n,Q^2)\}_{"LO"}\\
&+&{\alpha(Q^2)\over 2\pi}\delta C^G_n\Delta G(n,Q^2)_{"LO"}~,
\label{NLOmomg1}
\end{eqnarray}\\
\begin{equation}
M_n^{d}(Q^2)={1\over 2}[M_n^{(p)}(Q^2)+M_n^{(n)}(Q^2)](1-1.5\omega_D)~.
\label{momde}
\end{equation}
where $~\Delta q_3,~\Delta q_8~$ and $~\Delta \Sigma~$ are the moments 
of the flavour non-singlet and singlet combinations (\ref{a3x}), 
(\ref{a8x}) and (\ref{dsig}), respectively,
while $~G(n,Q^2)~$ denotes the moments of the gluon density $~G(x,Q^2)$.
The subscript "LO" in (\ref{NLOmomg1}) means that the moments of
the corresponding densities satisfy the LO $Q^2$-evolution
equations, in which for the strong coupling constant
$~\alpha_s(Q^2)~$ the NLO approximation
\begin{equation}
{\alpha_s(Q^2)\over 4\pi}={1\over \beta_0 ln{Q^2/\Lambda^2}} - 
{\beta_1\over \beta_0^3}{lnln{Q^2/\Lambda^2}\over ln{Q^2/\Lambda^2}}
\label{als}
\end{equation}
is taken. In (\ref{momde}) for the probability of the deuteron to
be in D-state we have taken $~\omega_d = 0.05 \pm 0.01$, which
covers most of the published values \cite{dfactor}. 

All quantities - the anomalous dimensions $~\delta\gamma^n_{ij}~$
up to two-loop approximation and the moments of the coefficient 
functions $~\delta {C^q_n}~$ and $~\delta {C^G_n}~$
in one-loop approximation needed to derive the analytic "LO"
and NLO solutions for the moments of the parton densities, can be
found, for instance, in \cite{Vog}.\\

Unlike paper \cite{Vog} where the expressions for the moments of
$~g_1~$ and $~2xF_1~$ are numerically Mellin-inverted to
yield the structure functions in Bjorken $x$-space, we follow 
a method \cite{Parisi, Kriv} which presents the structure functions 
{\it analytically.} 
Having the NLO $~Q^2$-evolution of the moments (\ref{NLOmomg1}) and 
(\ref{momde}) we can write the structure 
function $~g_1(x,Q^2)_{NLO}~$ in the form:
\begin{equation}
xg_{1}^{N}(x,Q^2)=x^{\a}(1-x)^{\beta}\sum_{n=0}^{N_{max}}\Theta_n^
{\a ,\beta}(x)\sum_{j=0}^{n}c_{j}^{(n)}{(\a ,\beta )}
M_{j+2}^{NS} \left ( Q^{2}\right ),  
\label{Jacobi}
\end{equation}\\
where $~\Theta^{\alpha \beta}_{n}(x)~$ is a set of Jacobi polynomials 
and $~c^{(n)}_{j}(\alpha,\beta)~$ are coefficients of the series of
$~\Theta^{\alpha,\beta}_{n}(x)~$ in powers in x:
\begin{equation}
\Theta_{n} ^{\a , \beta}(x)=
\sum_{j=0}^{n}c_{j}^{(n)}{(\a ,\beta )}x^j .
\label{e9}
\end{equation}

$N_{max},~ \alpha~$ and $~\beta~$ have to be chosen so as to
achieve the fastest convergence of the series on the R.H.S.
of Eq. (\ref{Jacobi}) and to reconstruct $xg_1$ with
the required accuracy. We use $~\alpha = 0.7~,
~\beta = 3.0~$ and $~N_{max} = 8~$. These values guarantee an accuracy 
better than 2.5\% in the experimental $~x~$ range:$~0.01\leq x < 0.8~$
and better than 5\% for smaller $x$.

The same method has been applied to calculate the unpolarized
structure functions $~2xF_1^{N}(x,Q^2)_{NLO}~$ from their moments 
(\ref{momF1}). Following the results of Ref. \cite{Kriv} we use
for the quantities $N_{max},~ \alpha~$ and $~\beta~$ in this case:
$$N_{max}=12,~~~~\alpha=-0.85,~~~~\beta=3.0~$$
in order to guarantee an accuracy better than $~10^{-3}~$ in 
the $~x~$ range
mentioned above. In the present calculations of $~2xF_1~$ the
MRS(A\'{}) parametrization for the input unpolarized
parton densities has been used.\\

As already mentioned in the introduction all
calculations are performed in $~\overline {MS}~$ scheme. In this
scheme the first moment of the Wilson coefficient function  
$~\delta {C^G_1}=0~$, so the axial anomaly contribution of
$~\Delta G(1,Q^2)~$ to the first moment $~M^N_1(Q^2)\equiv 
\Gamma_1^N(Q^2)~$ of $~g_1(x,Q^2)~$ vanishes, {\it i.e.} the
singlet axial charge $~a_0(Q^2)~$ coincides with $~\Delta \Sigma
(1,Q^2)~$ (twice the total helicity) carried by the quarks in the 
nucleon. In
the present study we use the $~\overline {MS}~$ scheme because of
our choice of the input parton distributions (see, Eqs. (\ref{Brunpol}) 
and (\ref{classic})).
A NLO analysis of the polarized DIS data in a renormalization scheme,
which allows for an anomalous gluonic contribution $~\Delta G(1,Q^2)~$
to the Ellis-Jaffe sum rule \cite{ElJa} is presented in \cite{anomdg}.

The value of $~\Lambda_{\overline{MS}}~$ is taken to be $~\Lambda_
{\overline{MS}}(n_f=3)=284~MeV.$ This value corresponds to
$~\Lambda_{\overline{MS}}(n_f=4)=231~MeV~$ \cite{MRS} according
to the requirement $~\alpha_s~$ be continuous across each threshold
$~Q^2\geq m_{q}^2$. The contributions of charmed quarks to 
$~g_1(x,Q^2)~$ is assumed to be negligible at present energies 
\cite{GRV} and will not be considered in our analysis.

The last step before fitting the theoretical predictions for $~A_1(x,Q^2)~$
to the data is to choose the input polarized parton densities at
some fixed value of $~Q^2=Q^2_0~$, evolve them to $~Q^2~$ and
then put them into (\ref{Jacobi}).\\

{\bf 4. Models for the input parton distributions}
\vskip 4mm
 
We have studies two classes of models:

(a) the model of Brodsky, Burkardt and Schmidt (BBS) \cite{BBS}, which
directly parameterizes the parton densities $~q_{\pm}(x)~$ and
which respects the perturbative QCD counting rules exactly

(b) an example of parametrization of the form $~\Delta q(x)=f(x)q(x)~$ 
where $~q(x)~$ are the unpolarized parton densities of Martin,
Roberts and Stirling MRS(A\'{}) \cite{MRS} and in which the counting
rules are somewhat relaxed.\\

(a) {\it The BBS counting rule model}\\

It is well known that the valence Fock states with the minimum
numbers of constituents dominate in determining the behaviour of the
unpolarized valence distributions $~u_{v}(x),~d_{v}(x)~$ in the region
$x\to 1~$ and this leads \cite{unpolCR}, {\it via} perturbative QCD, to
the prediction
\begin{equation}
u_{v}(x),~d_{v}(x)\sim (1-x)^3~~~~~ (x\to 1).
\label{CRuvdv}
\end{equation}  
The same arguments applied to the polarized case suggest 
\cite{Far,BBS} that
\begin{equation}
q_{v_{+}}(x)\sim (1-x)^3,~~~~~~q_{v_{-}}(x)\sim (1-x)^5~~~~~(x\to 1)
\label{CRqpqm}
\end{equation}
implying, {\it via} (\ref{CRuvdv}), that
\begin{equation}
{\Delta q_{v}(x)\over q_{v}(x)}\to 1~~~~~~(x\to 1)~.
\label{CRdqdivq}
\end{equation}

Simple perturbative arguments based upon the splitting functions for
$~q\to qG~$ and $~G\to q\bar{q}~$ can then be used to predict the
behaviour of the gluon and sea densities generated from the
valence quarks in the region $~x\to 1~$, namely \cite{Close,BBS}
\begin{equation}
G(x)\sim (1-x)^4,~~~~~~{\Delta G(x)\over G(x)}\to 1~~~~~(x\to 1)
\label{CRG}
\end{equation}
and
\begin{equation}
\bar{q}(x)\sim (1-x)^5,~~~~~~{\Delta \bar{q}\over \bar{q}}\to
1~~~~~(x\to 1)~.
\label{CRsea}
\end{equation}

Unfortunately these simple sum rules are not compatible with the 
evolution
equations. If they hold at some $~Q^2_0~$ then the power of
$~(1-x)~$ involved will grow like $~lnlnQ^2~$ at large $~Q^2~$.
It is therefore not clear to what extent the above rules should
hold at some $~Q^2_0~$ at which one chooses to parameterize the
parton densities. Presumably they should be more accurate at
momentum scales $~Q_0\sim \Lambda_{QCD}$, so that for the typical
values of $~Q^2_0~$ utilized in the data analyses, $~1\leq Q^2_0
\leq 4~GeV^2$, one might expect somewhat higher powers of $~(1-x)~$
to appear. Indeed, analyses of the unpolarized data yield typically,
at $~Q^2_0=4~GeV^2$,
\begin{equation}
u_v(x)\sim (1-x)^{(3~-~4)}~,~~~~~~~d_v(x)\sim (1-x)^{(4~-~5)}
\label{CRunpol}
\end{equation}
somewhat at variance with (\ref{CRuvdv}) and with the faster
decrease of $~d_v(x)~$ being related to the experimentally established 
behaviour of $~F_2^n(x)/F^p_2(x)~$ as $~x\to 1$.

The BBS model is interesting because it asks how well the data
can be fitted if the exact conditions (\ref{CRuvdv} - \ref{CRsea}) 
are imposed.

Concerning the behaviour as $~x\to 0$, perturbative arguments
based on the splitting functions predict \cite{BBS} that the
gluons and antiquarks
\begin{equation}
{\Delta \bar{q}(x)\over \bar{q}(x)}~,~~{\Delta G(x)\over G(x)}\to
const.x~~~~~(x\to 0)~.  
\label{Regge}
\end{equation}
This behaviour is consistent with Regge-type arguments.

For the gluons BBS use wave-function arguments to suggest that
the {\it const} in (\ref{Regge}) is approximately equal to 1 and
their gluon density automatically satisfies this constraint. For
the sea the {\it const} is determined by the fit to the data.\\

In the original BBS paper \cite{BBS} an analysis of the polarized
data then available and a fit to the MRS(D0\'{}) parametrization of
the unpolarized data were performed, but without taking account of 
the corrections from QCD evolution.
As on the one hand, there is now much more polarized data
available, especially at smaller $x$, and the MRS(D0\'{}) fit has
been superseded by other parametrizations which fit the
unpolarized low-$x$ data far better, and on the other, the NLO
calculations in QCD have been completed, we have felt it
important to examine to what extent the original claims of the
BBS model remain valid.\\ 

The input helicity-dependent parton densities at $~Q^2_0~$
in the BBS model have the following form: 
\begin{eqnarray}
\nonumber
x(\Delta u(x)+\Delta \bar{u}(x))&=&x^{1-\alpha_q}(1-x)^3
[A_u+B_u(1-x)-C_u(1-x)^2-D_u(1-x)^3]~,\\
\nonumber
x(\Delta d(x)+\Delta \bar{d}(x))&=&x^{1-\alpha_q}(1-x)^3
[A_d+B_d(1-x)-C_d(1-x)^2-D_d(1-x)^3]~,\\
\nonumber
x(\Delta s(x)+\Delta \bar{s}(x))&=&x^{1-\alpha_q}(1-x)^5
[A_s+B_s(1-x)-C_s(1-x)^2-D_s(1-x)^3]~,\\
x\Delta G(x)&=&x^{1-\alpha_g}(1-x)^4[A_g+B_g(1-x)][1-(1-x)^2]~,
\label{Brpol}
\end{eqnarray}
while the unpolarized parton densities at $~Q^2_0~$ are given as
\begin{eqnarray}
\nonumber
x(u(x)+ \bar{u}(x))&=&x^{1-\alpha_q}(1-x)^3[A_u+B_u(1-x)+
C_u(1-x)^2+D_u(1-x)^3]~,\\
\nonumber
x(d(x)+ \bar{d}(x))&=&x^{1-\alpha_q}(1-x)^3[A_d+B_d(1-x)+
C_d(1-x)^2+D_d(1-x)^3]~,\\
\nonumber
x(s(x)+ \bar{s}(x))&=&x^{1-\alpha_q}(1-x)^5[A_s+B_s(1-x)+
C_s(1-x)^2+D_s(1-x)^3]~,\\
xG(x)&=&x^{1-\alpha_g}(1-x)^4[A_g+B_g(1-x)][1+(1-x)^2]~.
\label{Brunpol}
\end{eqnarray}

The constraints 
\begin{equation}
A_q+B_q=C_q+D_q
\label{constr}
\end{equation}
are imposed on the constants $~A_q,~B_q,~C_q~$ and $~D_q~$ in
(\ref{Brpol}) and (\ref{Brunpol}) in order to ensure the
convergence of the helicity-dependent sum rules (\ref{BSR}) and
(\ref{BHSR}). Thus in the BBS
model the Regge behaviour of the polarized quark densities
$~\Delta q\sim x^{-\alpha_R}~$ is automatically one unit less than the
unpolarized intercept $\alpha_q:\alpha_R=\alpha_q-1$. Isospin
symmetry at low $~x~$  requires
\begin{equation}
A_u+B_u+C_u+D_u = A_d+B_d+C_d+D_d~.
\label{isospin}
\end{equation}

If in addition to (\ref{constr}) and (\ref{isospin}) the helicity-
dependent sum rules (\ref{BSR}) and (\ref{BHSR}) and the energy-momentum
sum rule for the unpolarized densities (\ref{Brunpol}) are taken
into account, the number of the unknown parameters associated
with the input polarized densities is reduced to $~N=16-7=9$. These
free parameters - we have chosen 
\begin{equation}
\{B_u,~C_u,~D_u,~C_d,~C_s,~D_s,~B_g,~\alpha_q,~\alpha_g\}~, 
\label{freepar}
\end{equation}
are determined by a {\it simultaneous}
fit of the theoretical predictions (\ref{NLOas}) and the BBS {\it
unpolarized} parton densities (\ref{Brunpol}) to the world 
$~A^N_1(x,Q^2)~$ data and the MRS(A\'{}) set of unpolarized parton
densities, respectively.  
We recall that the MRS(A\'{}) densities at next-to-leading
order are parameterized at $~Q^2_0=4~GeV^2~$ and are well
determined from the wide set of unpolarized experiments. Note
that the choice of unpolarized densities is not crucial for the
present analysis.\\

(b) {\it Models based on the unpolarized parton densities}\\

We have also analyzed the polarized DIS data using the following
expressions for the input polarized parton distributions at $~Q^2_0$  
\begin{eqnarray}
\nonumber
x\Delta u_v(x,Q^2_0)&=&\eta_u A_ux^{a_u}xu_v(x,Q^2_0)~,\\
\nonumber
x\Delta d_v(x,Q^2_0)&=&\eta_d A_dx^{a_d}xd_v(x,Q^2_0)~,\\
\nonumber
x\Delta Sea(x,Q^2_0)&=&\eta_s A_sx^{a_s}xSea(x,Q^2_0)~,\\
x\Delta G(x,Q^2_0)&=&\eta_g A_gx^{a_g}(1-x)^{b_g}xG(x,Q^2_0)
\label{classic}
\end{eqnarray}
where on R.H.S. of (\ref{classic}) we use the MRS(A\'{}) 
unpolarized densities. The normalization factors $~A_f~$
are determined in such a way as to ensure that the first moments of
the polarized densities are given by $~\eta_{f}$. Since the
present polarized experiments do not allow for a flavour decomposition 
of the sea, SU(3) symmetry of the sea-quark densities is assumed 
\begin{equation}
\Delta \bar{u}(x,Q^2_0)= \Delta \bar{d}(x,Q^2_0)= \Delta \bar{s}(x,Q^2_0) 
=\Delta \bar{q}(x,Q^2_0)~.
\label{seasym}
\end{equation}
Following this assumption $~\Delta Sea(x,Q^2_0)=6\Delta \bar{q}
(x,Q^2_0)~$ and $~\eta_s=6\eta_{\bar{q}}~$, where $~\eta_{\bar{q}}~$
is the first moment of $~\Delta \bar{q}~$. 
Note also that the very small
charm contribution to the unpolarized sea on R.H.S. of 
(\ref{classic}) is ignored in our analysis.

In this approach, the first moments of the valence quark densities
$~\eta_u~$ and $~\eta_d~$ are obtained directly from the sum rules
(\ref{BSRa}) and (\ref{BHSRa})
\begin{equation}
\eta_u=0.918,~~~~~~~\eta_d=-0.339~.
\label{etaud}
\end{equation}

The first moment of the polarized sea $~\eta_s~$ is fixed from the 
measured value of $~\Gamma_1$. More details are discussed in the next
section. The rest of the parameters
\begin{equation}
\{a_u,~a_d,~a_s~,\eta_g~,a_g~,b_g\}~,
\label{claspar}
\end{equation}
have to be determined from the fit to the $~A_1^N(x,Q^2)~$ data.

Finally it should be noted that using a set of polarized parton
densities like (\ref{classic}), $~\Delta d_v(x,Q^2_0)~$ and   
$~\Delta \bar{q}(x,Q^2_0)~$ will turn out to be negative in the 
whole $x$ region in contrast to the BBS model where these quantities 
become positive at large $x$.\\

{\bf 5. Results of Analysis}
\vskip 4mm

In this section we present the results of our fits to the world 
$~A_1^N(x,Q^2)~$ data: EMC proton data \cite{EMC}, SLAC E142
neutron data \cite{SLACn}, SLAC E143 proton and deuteron data [5 - 7],
SMC proton data \cite{SMCp97} and the SMC deuteron data \cite{SMCd97}
which are combined data from the 1992 \cite{SMCd93}, 1994 \cite{SMCd95}
and 1995 runs. The data used (203 experimental points) cover the
following kinematic region:  
\begin{equation}
0.004< x < 0.75,~~~~~~1< Q^2< 72~GeV^2~.
\label{kinreg}
\end{equation}

We have chosen $~Q^2_0=4~GeV^2$. In all fits only statistical
errors are taken into account. "Higher twist" corrections are not
included in the present study. As already discussed
above, in the approach used their effect is expected to be 
negligible.\\

(a) {\it BBS model}\\

A comparison of our results (solid curves) with the data on $~A_1^N
(x,Q^2)~$ is shown in Figs. 2a - 2f. Our NLO results for
$~g_1^N(x,Q^2)~$ are illustrated in Fig. 7.
The minimum of the functional $~\chi^2~$ is
achieved at $~\chi^2=232.7~$ and $~\chi^2/DOF=232.7/194=1.20$.
The values of the free input parton parameters (\ref{freepar})
corresponding to this $~\chi^2~$ value are
\begin{eqnarray}
\nonumber
&&B_u=-3.010 \pm 0.156,~~~~C_u=2.143 \pm 0.137,~~~~D_u=-2.065\pm
0.148~,\\
\nonumber
&&C_d=~~1.689 \pm 0.227,~~~~C_s=0.334 \pm 0.044,~~~~D_s=-0.292\pm 
0.042~,\\
&&B_g=-0.339 \pm 0.454,~~~~\alpha_q=1.313 \pm 0.056,~~~~\alpha_g=
1.233\pm 0.073~.
\label{valpar}
\end{eqnarray}

The rest of the parameters $~A_u,~A_d,~B_d,~D_d,~A_s,~B_s~$ and
$~A_g~$ are determined by the constraints (\ref{constr}) and 
(\ref{isospin}),
the sum rules (\ref{BSR}) and (\ref{BHSR}) and the momentum-energy
sum rule, giving
\begin{eqnarray}
\nonumber
A_u=3.088,&&~~A_d=0.343,~~~~~B_d=-0.265,~~~~~D_d=-1.610~,\\
          &&A_s=0.001,~~~~B_s=0.041,~~~~A_g=1.019~.
\label{SRpar}
\end{eqnarray}

It is seen from Figs. 2a - 2f  that the NLO QCD predictions using 
the BBS model for the input parton distributions are in a good 
agreement with the presently available data on $~A_1^N(x,Q^2)~$, as
well as with the corresponding $~g_1^N(x,Q^2)$ data (see Fig. 7).

For the first moments of the polarized flavour and flavour-singlet 
distributions at $~Q^2_0=4~GeV^2~$ we obtain
\begin{equation}
\Delta u+\Delta\bar{u}=0.839,~~~~\Delta d+\Delta\bar{d}=-0.405,
~~~~2\Delta\bar{s}=-0.079~,
\label{delq1}
\end{equation}
and
\begin{equation}
\Delta \Sigma=a_0=0.342~.
\label{delsig1}
\end{equation}

These values yield for the quantity $~\Gamma^N_1(Q^2)~$, the
first moment of $~g_1^N~$,
\begin{equation}
\Gamma_1^p(5~GeV^2)=0.146,~~~\Gamma_1^n(5~GeV^2)=-0.047,~~~
\Gamma_1^d(5~GeV^2)=0.046~,
\label{Gam1}
\end{equation}
which are in a good agreement with their experimental values 
\begin{eqnarray}
\nonumber
&&\Gamma_1^p(5~GeV^2)=0.141\pm 0.011~~~~~~(All~proton~data)~,\\
&&\Gamma_1^d(5~GeV^2)=0.039\pm 0.004~~~~~~(All~deuteron~data)~.
\label{Gam1exp}
\end{eqnarray}

The experimental values of $~\Gamma^p_1~$ and $~\Gamma^d_1~$
in (\ref{Gam1exp}) have been determined in \cite{SMCp97} and
\cite{SMCd97} by combining SMC and SLAC E143 results on $~A^p_1~$
and $~A^d_1$, respectively.

Our result for the axial charge $~a_0~$
\begin{equation}
a_0(5~GeV^2)=\Delta \Sigma(5~GeV^2)=0.341
\label{axial}
\end{equation}
is also in a good agreement with its experimental value 
$~a_0(5~GeV^2)=0.29\pm 0.06~$  recently determined \cite{SMCp97}
from the combined analysis of all proton, neutron and deuteron data.

We obtain for the small $x$ behaviour of the input sea quark and
gluon distributions
\begin{equation}
x\Delta\bar{q}(x)\sim x^{0.69},~~~~~~x\Delta G(x)\sim x^{0.77}~.
\label{smallx}
\end{equation}
This result confirms that of Gehrmann and Stirling 
\cite{GSt} who used in their NLO analysis the same
renormalization scheme and the same $~Q^2_0~$, but different
input parton densities.

It is well known that in the unpolarized case the gluon distribution
can not be well determined by the fit to the data on nucleon structure
functions alone. The situation in the polarized DIS is even worse. 
In particular, the fact that the parameter $~B_g~$ in (\ref{valpar}) is
not well determined reflects this uncertainty in the extraction of 
$~\Delta G(x,Q^2)~$ from the data. We obtain for the first moment of 
$~\Delta G(x,Q^2):~\Delta G_1(4~GeV^2)=0.447~$, which does not
coincide with the mean values of this quantity given in most
of the theoretical analyses. This result for $~\Delta G_1(Q^2_0)~$
is not surprising. It should be noted that in the BBS model one
can show that $~\Delta G_1(Q^2_0)~$ is constrained by
\begin{equation}
\Delta G_1(Q^2_0) < 2G_2(Q^2_0) - G_3(Q^2_0)~,
\label{G1SR}
\end{equation}
where $~G_2(Q^2_0)~$ and $~G_3(Q^2_0)~$ are the second and third
moments of the unpolarized densities. Bearing in mind that
from the unpolarized data $~G_2(Q^2_0)\sim 0.45~$ one has that
$~\Delta G_1(Q^2_0)~$ should be smaller than 0.9.
This fact could be critical for the validity of the BBS model if
more precise data are available in the future.\\

The BBS polarized $~x(\Delta f(x)+\Delta\bar{f}(x))~$  distributions 
at $~Q^2_0=4~GeV^2~$ are shown in Fig. 3. In Fig. 4 we compare our 
results for the BBS {\it unpolarized} $~x(f(x)+\bar{f}(x))~$ input 
densities (solid curves) with the MRS(A\'{}) set of unpolarized 
parton distributions (dashed curves) at the same $~Q^2_0=4~GeV^2$. 
The difference between the BBS unpolarized input parton densities
and the MRS(A\'{}) ones is somewhat greater than the
usual difference between the various sets of unpolarized parton
distributions used in the
literature. The difference for $~x(u+\bar{u})~$ and $~x(d+\bar{d})~$
quarks increases up to 20\% depending on $x$, while 
{\it e.g.}, for MRS(D0\'{}) and MRS(A\'{}) set of input valence quark 
densities it is typically 6-8\% in the $x$ range:
$0.01\leq x\leq 0.50~$ and amounts to 15-20\% for the range: 
$0.5<x\leq 0.7$.
As one can see from Fig. 4 the agreement for the sea
quarks and the gluons is better, of the same quality as for the
well known different sets of unpolarized input partons.\\

(b) {\it Models based on the unpolarized parton densities}\\

Let us continue now with discussion of our results of the fit to the 
data using for the input polarized parton densities the set 
(\ref{classic}). Such a set of input partons had been used in \cite{Vog}, 
but starting at very small $~Q_0^2=0.34~GeV^2$.

Unlike the BBS model the data do not allow to determine $~\eta_s$, 
the first moment of
the polarized sea, in proper way if the parametrization (\ref{classic})
is used. In order words, if $~\eta_s~$ is taken to be a free parameter,
its value determined from the fit to $~A_1^N(x,Q^2)~$ data, does
not agree with the experimental value of $~\eta_s/3=-0.10\pm
0.02~$ \cite{SMCp97}. That is why we fix $~\eta_s~$ from the
measured value of $~\Gamma_1^p(5~GeV^2)=0.141\pm 0.011~$ to be
$~\eta_s=-0.290~$ at $~Q_0^2=4~GeV^2$.

As already mentioned above it is impossible to determine
accurately the form of the polarized gluon density from these
data alone and therefore additional constraints have to be applied 
to the gluonic input parameters. As in \cite{GSt} we have used the
assumption $~a_g=a_s~$ which defines the behaviour of $~\Delta 
G(x,Q^2)~$ at small $x$. It turns out that even in that
case it is not possible to determine by the fit the value of
the parameter $~b_g~$ in (\ref{classic}) which controls the
behaviour of $~\Delta G(x,Q^2)~$ at large $x$. For that reason
the fit to the data was performed at different fixed values of 
$~b_g$ in the range $~0\leq b_g \leq 7$. Further, if $~a_d~$ is
left as a free parameter, its best-fit value is negative
independently of the value of $~b_g~$ in the above range. This
fact, presumably reflects the circumstance discussed in Section 2,
namely that from the data in the small $x$ region 
$~g_1^p-g_1^n > g_1^p + g_1^n~$, which probably induces such a 
behaviour in $~\Delta d_v(x,Q^2_0)~$. For this negative value 
of $~a_d:a_d\sim -0.13~$ 
$${\vert {\Delta d_v(x)}\vert\over d_v(x)} = 0.339A_dx^{a_d}$$
is greater than 1 at very small $x:~x<0.3.10^{-6}~$ and the 
positivity condition (\ref{posit}) is broken. In addition, the 
corresponding value of $~a_s~$ determined by this fit is such 
that the positivity condition (\ref{posit}) for $~\Delta Sea(x)~$ 
is violated too, but at large $x:~x > 0.80$. That is why we have 
taken $~a_d=0~$, the smallest value of $~a_d~$ which guarantees 
positivity for $~\Delta d_v$ at all $x$. The change of $~\chi^2~$ 
from $~\chi^2(a_d<~0)\cong 215~$ to $~\chi^2(a_d=0)
\cong 220~$ is negligible.

The results of the fit to $~A_1^N(x,Q^2)~$ data are presented in
Table 1. 
\vskip 0.6 cm
\begin{center}
\begin{tabular}{cl}
&{\bf Table 1.} The results of the NLO QCD fit
to the world $~A_1^N~$ data using the set\\ 
&(\ref{classic}) for the input polarized partons ($~a_d=0$).
\end{tabular}
\vskip 0.6 cm
\begin{tabular}{|c|c|c|c|c|c|c|} \hline
             & $b_g=0$  & $b_g=5$  & $b_g=7$ \\ \hline
 $\chi^2/DOF$& 221.2/200        &~220.5/200~        &~219.5/200~\\
 $a_u$       & 0.196~$\pm$~0.018&~0.178~$\pm$~0.027~&~0.169~$\pm$~0.025~\\
 $a_s$       & 0.892~$\pm$~0.085&~0.706~$\pm$~0.155~&~0.694~$\pm$~0.112~\\
 $\eta_g$    & -0.14~$\pm$~0.26 &~1.07 ~$\pm$~0.70 ~&~1.29
~$\pm$~0.53 ~\\ \hline
\end{tabular}
\end{center}
\vskip 0.6 cm

Although only three free parameters
$~\{a_u,~a_s,~\eta_g\}~$ have been used a very good description
of the data is achieved. This is illustrated in Figs. 2a - 2f
(dashed curves), which show the NLO description of the various 
$~A_1^N~$ measurements. In Fig. 7 we compare our NLO results for
$~g_1^N(x,Q^2)~$ with the SMC data. The input polarized parton 
distributions (\ref{classic}) are shown in Fig. 5.
The value of $~\chi^2/DOF~$ is 220.5/200 for $~b_g=5~$ and
practically does not depend on $~b_g~$ (see the Table). This
value is better than that one in the case of the BBS model.
The mean value of $~\eta_g~$, the first moment of the polarized 
gluon density, is sensitive to the value of $~b_g:\eta_g~$
increases with increasing $~b_g$. The values of $~\eta_g~$
at $~b_g\geq 5~$ are greater than 1 and are in agreement with those
determined by the NLO analysis in \cite{GSt}. The sets of 
polarized quark densities corresponding to gluons with $~b_g=5~$
and $~b_g=7~$ are practically the same.

We have found also that values for $~\eta_g~$ smaller than 1, 
and even a small negative value $~\eta_g = -0.14\pm 0.26~$
corresponding to $~b_g=0~$, are not excluded by the  data. 
In the last case, however,  
$${\vert {\Delta Sea(x)}\vert\over Sea(x)} = 1.12x^{0.892}$$
and positivity is violated in the large $x$ region:$~x>0.88$. \\

In Fig. 6 a comparison between the BBS input polarized
distributions (\ref{Brpol}) and the polarized densities (\ref{classic})
is shown. It is seen that while the distributions $~x(\Delta u
+ \Delta \bar{u})~$ and $~2x\Delta \bar{s}=2x\Delta \bar{q}~$ are 
very similar for these two parametrizations, the polarized
distributions $~x(\Delta d + \Delta \bar{d})~$ and $~x\Delta G~$
corresponding to (\ref{Brpol}) and (\ref{classic})
differ considerably, $~x(\Delta d + \Delta \bar{d})~$
for $~x>0.35~$ and $~x\Delta G~$ for all $x$.
The polarized gluon distribution enters $~g_1(x,Q^2)~$ at
next-to-leading order and bearing in mind the accuracy of the
present data these different gluon contributions can not be 
distinguished.
However, the difference between $~x(\Delta d + 
\Delta \bar{d})(x,Q^2_0)~$ leads to a 
considerably different behaviour of $~A_1^d(x,Q^2)~$ in the kinematic
region: $~x>0.35,~Q^2\sim 5-10~GeV^2$, and allows a better fit to
the SLAC E143 data on $~A_1^d(x,Q^2)~$ in this region in the case of
the parametrization (\ref{classic}) for the input polarized
parton densities. The difference for $~x(\Delta d + \Delta \bar{d})~$
at large $x$ is a consequence of the fact that the BBS
distributions are forced to satisfy (\ref{CRdqdivq}) as $x\to 1$.\\

{\bf 5. Conclusion}
\vskip 4mm

We have performed a next-to leading order QCD analysis 
($\overline{MS}$ scheme) of the  
world data on polarized deep inelastic lepton-nucleon scattering.
The QCD predictions have been confronted with the directly measured
virtual photon-nucleon asymmetry $~A_1^N(x,Q^2)~$ rather than
with the polarized structure function $~g_1^N(x,Q^2)~$ derived
from the data by
different additional procedures. Different parametrizations for
the input polarized parton densities have been examined: the Brodsky,
Burkardt and Schmidt model and a model based on unpolarized parton 
densities. In the BBS model the positivity constraints for the
unpolarized parton densities are automatically valid. In the
second one it is easy to control them.
In both cases a good fit to the data is achieved. The
description of the data is slightly better if for the input
polarized parton distributions the second parametrization is used.

A distinctive feature of the BBS model is that the
helicity-dependent input parton distributions $~f_{\pm}(x,Q^2_0)~$
(rather than $~\Delta f(x,Q^2_0)~$ and $~f(x,Q^2_0)~$) are
parameterized so that the polarized, as well as the unpolarized
data have to be fitted by the same parameters (\ref{freepar}).
This feature creates a serious challenge to the model.
We have shown that the BBS model is compatible with the present
DIS data although the agreement with the MRS(A\'{})
parametrization of the unpolarized parton distributions at
$~Q^2=4~GeV^2~$ is somewhat worse than the usual level of agreement
between the
different sets of the input unpolarized parton densities usually
used in the literature. Our result for the axial charge
$~a_0(5~GeV^2)=\Delta \Sigma(5~GeV^2)=0.341~$, obtained in the BBS 
model is in a good agreement with
its experimental value $~a_0(5~GeV^2)=0.29\pm 0.06~$  recently 
determined from a combined analysis of all proton, neutron and 
deuteron data, while using the second parametrization for the
input polarized densities this quantity is not well determined
from the fit to the data. This results from the fact that the sea
quark distributions are still largely undetermined if the 
second kind of parametrization is used. 

The present polarized data do not allow a precise determination
of the shape of the polarized gluon density. It follows from our fits 
that values of $~\Delta G_1(4~GeV^2)~$, the first moment of the 
polarized gluon density, both
greater and smaller than 1, are possible. Negative values of
$~\Delta G_1~$ are not excluded either.\\

Despite the great progress of the past few years it is clear that 
in order to test precisely the spin
properties of QCD more accurate DIS polarized data at fixed $x$ and
various $~Q^2~$ are needed. In addition, charged current data will
be very important
for a precise determination of the polarized parton densities and
especially, for a precise flavour decomposition of the polarized
quark sea. Finally, a direct measurement of $~\Delta G(x,Q^2)~$ in
processes such as $J/\psi$ production in lepton-hadron scattering
with a polarized beam will answer the important question about the
magnitude and the sign of $~\Delta G_1$.\\

\newpage
{\bf Acknowledgments}
\vskip 3mm

We are grateful to O. V. Teryaev for useful discussions and remarks.\\

This research was partly supported by a UK Royal Society Collaborative
Grant, by the Russian Fund for Fundamental Research Grant No 
95-02-04314a, by the INTAS Grant No 93-1180 and by the Bulgarian 
Science Foundation under Contract \mbox{Ph 510.}\\

\newpage
\noindent
{{\bf Figure Captions}}
\vskip 3mm
\noindent 
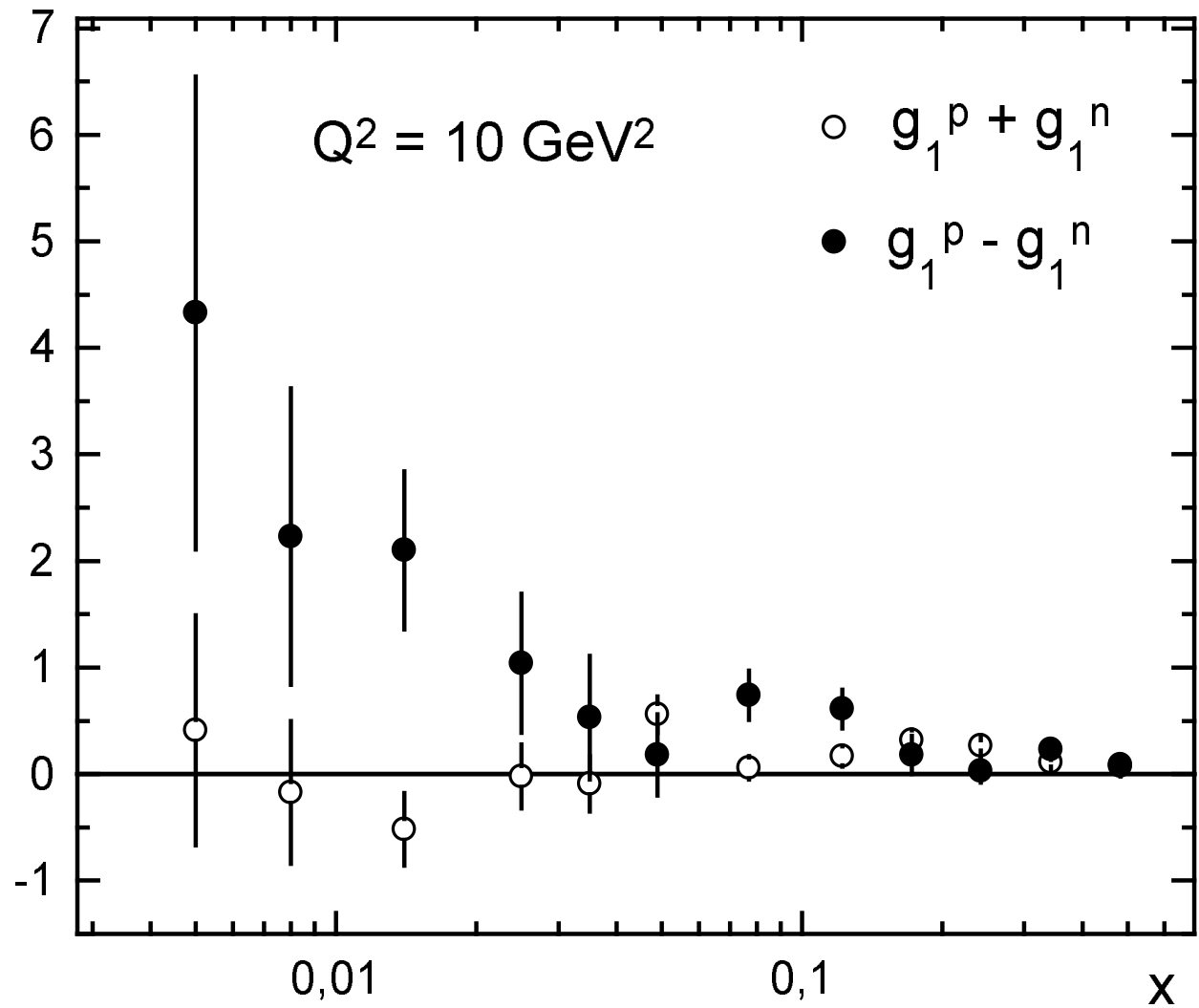
{\bf Fig. 1.} Comparison of the SMC data \cite{SMCp97,SMCd97} on
$~g_1^p\pm g_1^n~$ as function of $x$ at mean value $~Q^2=10
~GeV^2$.\\

\noindent 
{\bf Fig. 2a-2f.} Comparison of our NLO results for 
$~A_1^N(x,Q^2)~$
with the present data. The solid and dashed curves are the best
fits to the data corresponding to the BBS model (\ref{Brpol}) and 
parametrization (\ref{classic}) of input polarized parton 
distributions, respectively.\\  

\noindent 
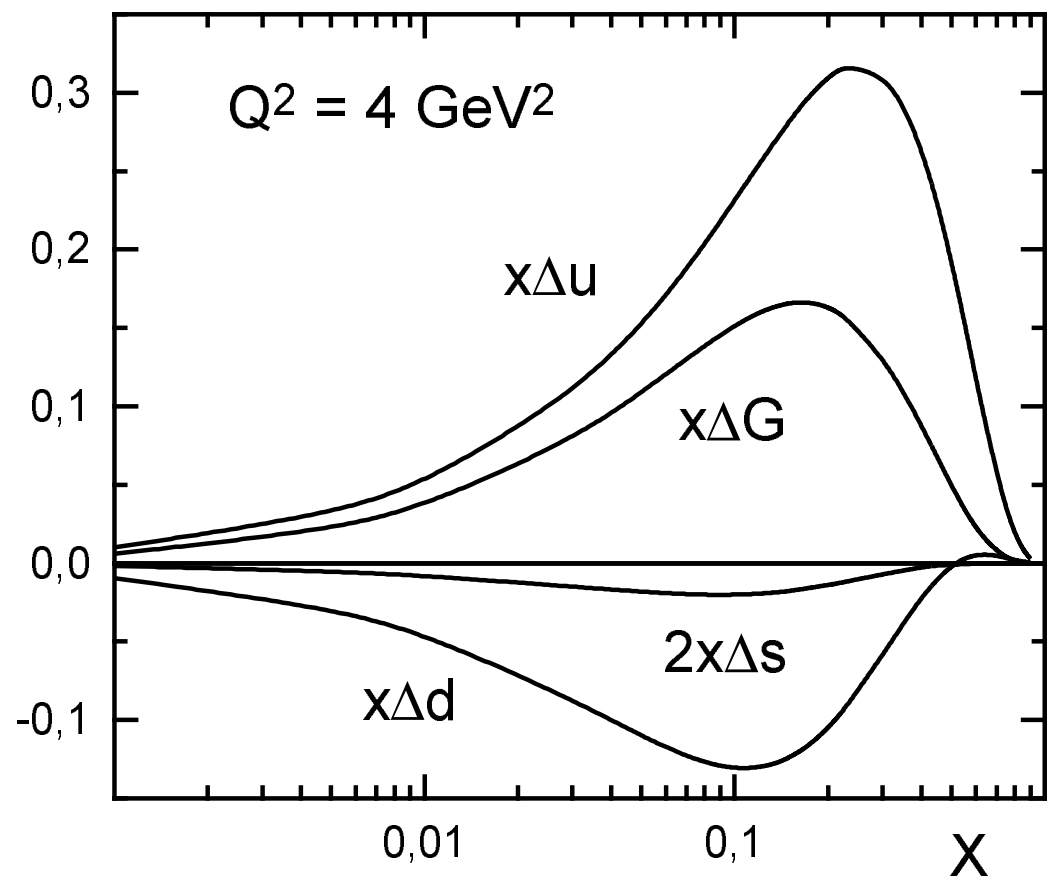
{\bf Fig. 3.} Next-to-leading order input polarized parton
distributions at $~Q^2=4~GeV^2~$ in the BBS model
($x\Delta u\equiv x\Delta u + x\Delta \bar{u}~$ and 
$x\Delta d\equiv x\Delta d + x\Delta \bar{d}$).\\

\noindent 
{\bf Fig. 4a-b.} Comparison between BBS {\it unpolarized} parton
distributions (solid curves) and MRS(A\'{}) set of unpolarized 
parton densities (dashed curves) at $~Q^2=4~GeV^2~
(xu\equiv xu+x\bar{u}~$ and $~xd\equiv xdu+x\bar{d}$).\\
 
\noindent 
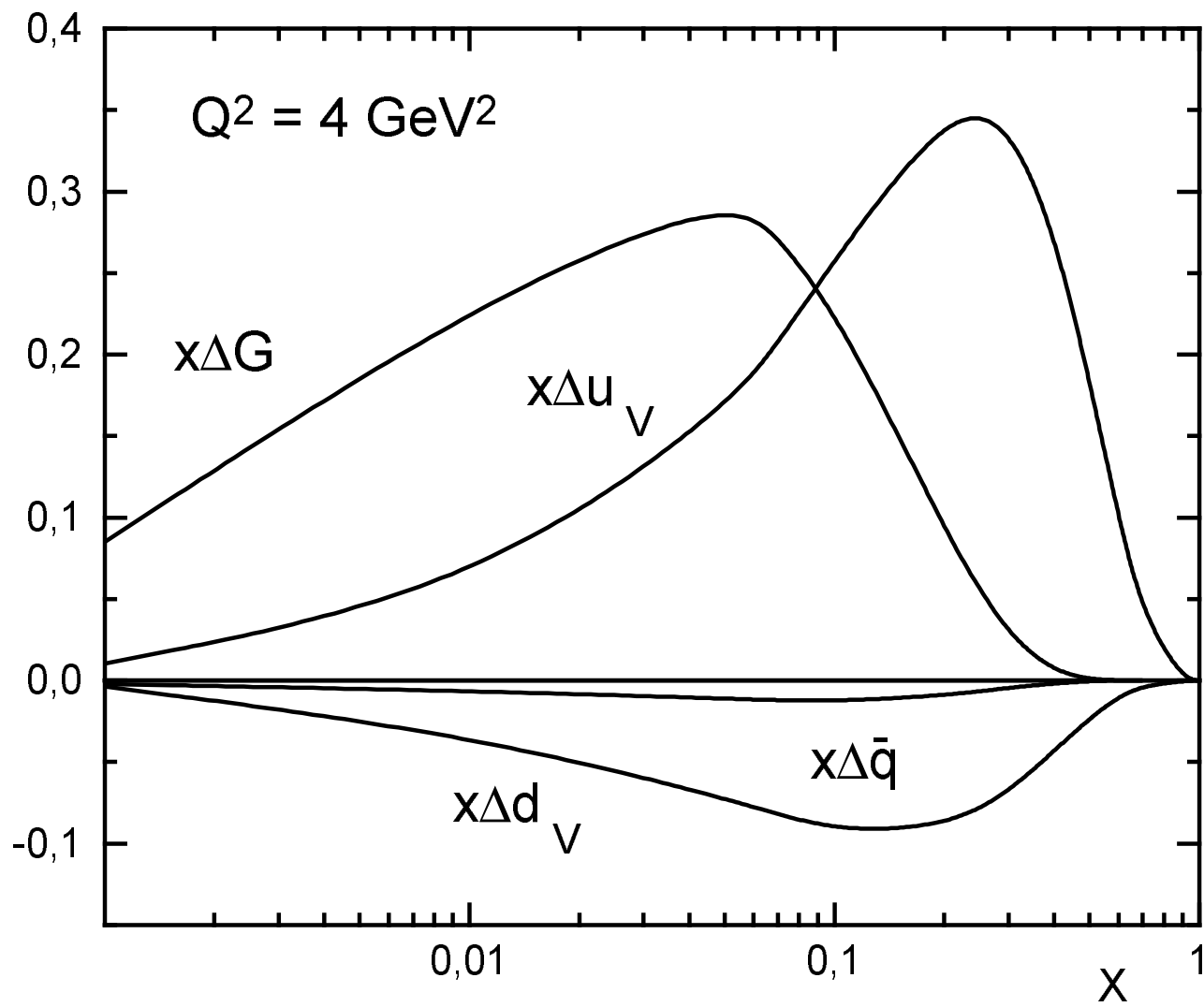
{\bf Fig. 5.} Next-to-leading order input polarized parton
distributions at $~Q^2=4~GeV^2~$ corresponding to parametrization
(\ref{classic}) in the text ($a_d=0,~b_g=7$).\\

\noindent 
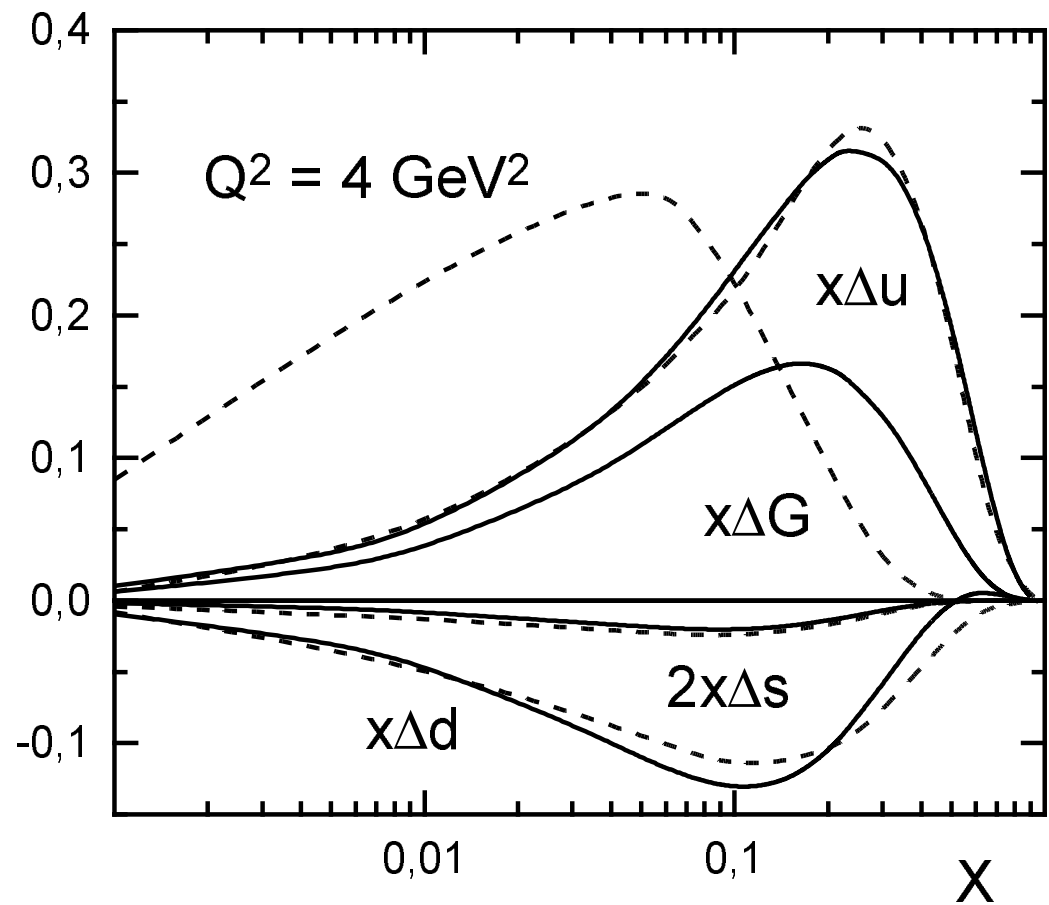
{\bf Fig. 6.} Comparison between the input polarized parton 
distributions in the BBS model (solid curves) and parametrization 
(\ref{classic}) (dashed curves). $~xu\equiv xu+x\bar{u},~xd\equiv 
xdu+x\bar{d}~$ and $~2x\Delta s=2x\Delta \bar{q}$.\\

\noindent 
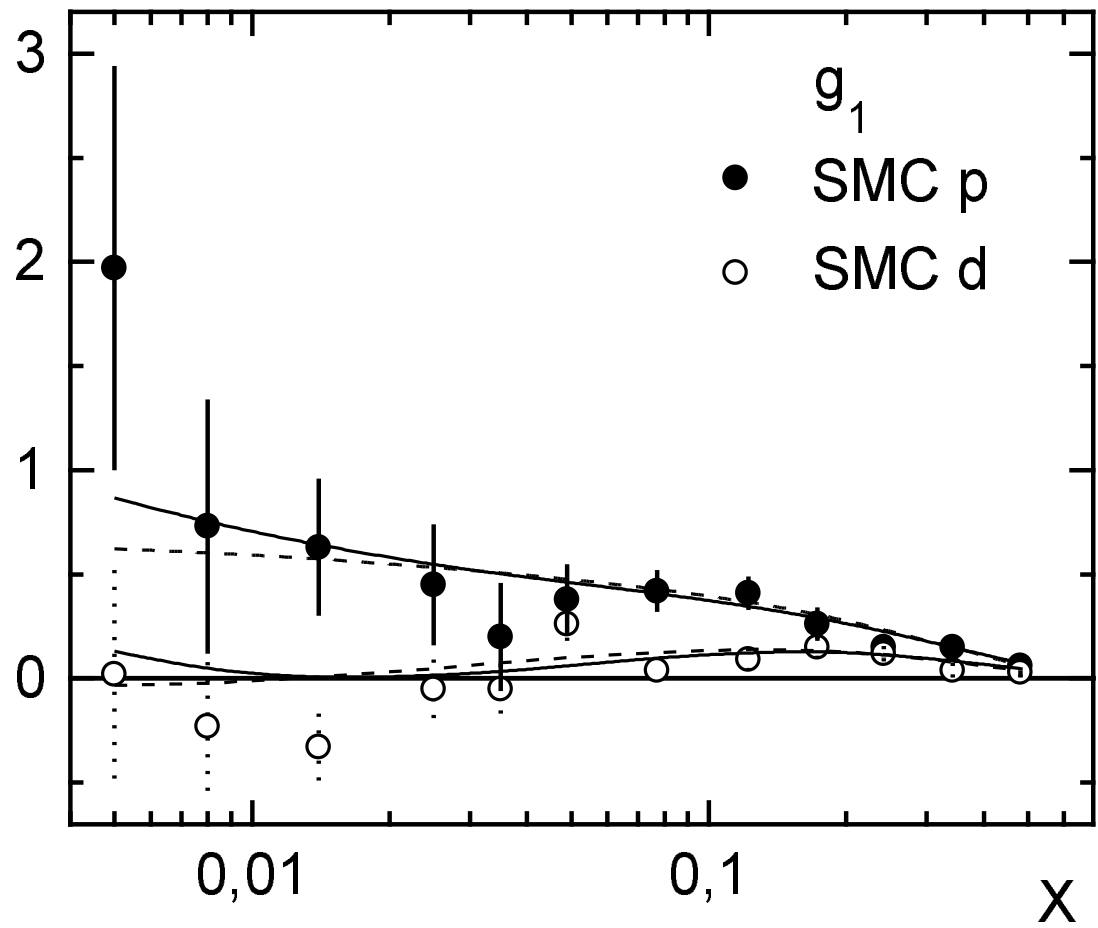
{\bf Fig. 7.} Comparison of our NLO results for $~g_1^N(x,Q^2)~$ 
with SMC data \cite{SMCp97, SMCd97} at the measured $~Q^2$.
The solid curves correspond to BBS model and dashed ones to
parametrization (\ref{classic}) of the input polarized parton
densities.


\newpage
\begin{thebibliography}{99}

\bibitem{Yale}
M. J. Alguard et al.,
\newblock Phys. Rev. Lett. {\bf 41} (1978) 70;\\
G. Baum et al., 
\newblock Phys. Rev. Lett. {\bf 51} (1983) 1135.

\bibitem{EMC}
EMC, J. Ashman et al.,
\newblock Phys. Lett. {\bf B206} (1988) 364;\\
Nucl. Phys. {\bf B328} (1989) 1.

\bibitem{ELA}
E. Leader and M. Anselmino,
\newblock Z. Phys. {\bf C41} (1988) 239.

\bibitem{SLACn}
SLAC E142, D. L. Anthony et al.,
\newblock Phys. Rev. Lett. {\bf 71} (1993) 959;

\bibitem{SLACp}
SLAC E143, K. Abe et al.,
\newblock Phys. Rev. Lett. {\bf 74} (1995) 346.

\bibitem{SLACd}
SLAC E143, K. Abe et al.,
\newblock Phys. Rev. Lett. {\bf 75} (1995) 25.

\bibitem{SLACpd}
SLAC E143, K. Abe et al.,
\newblock Phys. Lett. {\bf B364} (1995) 61.

\bibitem{SMCp94}
SMC, D. Adams et al,
\newblock Phys. Lett. {\bf B329} (1994) 399;\\
erratum {\it ibid} {\bf B339} (1994) 332.

\bibitem{SMCp97}
SMC, D. Adams et al, CERN preprint PPE/97-22, 1997, hep-ex/9702005.

\bibitem{SMCd93}
SMC, D. Adeva et al,
\newblock Phys. Lett. {\bf B302} (1993) 533.

\bibitem{SMCd95}
SMC, D. Adams et al,
\newblock Phys. Lett. {\bf B357} (1995) 248.

\bibitem{SMCd97}
SMC, D. Adams et al,
\newblock Phys. Lett. {\bf B396} (1997) 338.
%% CERN-PPE/97-08, 1997.

\bibitem{nlocor}
E. B. Zijlstra and W. L. van Neerven,
\newblock Nucl. Phys. {\bf B147} (1994) 61;\\
R. Mertig and W. L. van Neerven,
\newblock Z. Phys. {\bf C70} (1996) 637;\\
W. Vogelsang,
\newblock Phys. Rev. {\bf D54} (1996) 2023.

\bibitem{PDG}
Particle Data Group, L. Montanet et al,
\newblock Phys. Rev. {\bf D50} (1994) 1173;\\
F. E. Close and R. G. Roberts,
\newblock Phys. Lett. {\bf B313} (1993) 165.

\bibitem{Far}
G. R. Farrar and D. R. Jackson,
\newblock Phys. Rev. Lett. {\bf 35} (1975) 1416.

\bibitem{Close}
F. E. Close and D. Sivers,
\newblock Phys. Rev. Lett. {\bf 39} (1977) 1116.

\bibitem{BBS}
S. J. Brodsky, M. Burkardt and I. Schmidt,
\newblock Nucl. Phys. {\bf B441} (1995) 197.

\bibitem{MRS}
A.D. Martin, R.G. Roberts and W.J. Stirling,
\newblock Phys. Lett. {\bf B354} (1995) 155.

\bibitem{QCDsmallx}
R. D. Ball, S. Forte and G. Ridolfi,
\newblock Nucl. Phys. {\bf B444} (1995) 287.

\bibitem{Ryskin}
J. Batrels, B. I. Ermolaev and M. G. Ryskin,
\newblock Z. Phys. {\bf C70} (1996) 273;\\ 
DESY preprint 96-025, hep-ph/9603204.

\bibitem{NMC}
NMC, P. Amaudruz et al.,
\newblock Phys. Lett. {\bf B295} (1992) 159;\\
M. Arneodo et al.,
\newblock Phys. Lett. {\bf B364} (1995) 107.

\bibitem{SLAC}
L. W. Whitlow et al.,
\newblock Phys. Lett. {\bf B250} (1990) 193.

\bibitem{Vog}
M. Gl\"{u}ck, E. Reya, M. Stratmann and W. Vogelsang,
\newblock Phys. Rev. {\bf D53} (1996) 4775.

\bibitem{Stam}
D. B. Stamenov,
\newblock Mod. Phys. Lett. {\bf A10} (1995) 2029.

\bibitem{dfactor}
W. Buck and F. Gross,
\newblock Phys. Rev. {\bf D20} (1979) 2361;\\
M. Z. Zulhof and J. A. Tjon, 
\newblock Phys. Rev. {\bf C22} (1980) 2369;\\
M. Lacombe et al.,
\newblock Phys. Rev. {\bf C21} (1980) 861;\\
R. Machleidt et al.,
\newblock Phys. Rep. {\bf 149} (1987) 1.

\bibitem{Parisi}
G. Parisi and N. Sourlas,
\newblock Nucl. Phys. {\bf B151} (1979) 421;\\ 
I. S. Barker, C. B. Langensiepen and G. Shaw,
\newblock Nucl. Phys. {\bf B186} (1981) 61.

\bibitem{Kriv}
V. G. Krivokhizhin et al.,
\newblock Z. Phys. {\bf C36} (1987) 51;
{\it ibid} {\bf C48} (1990) 347.

\bibitem{ElJa}
J. Ellis and R. L. Jaffe,
\newblock Phys. Rev. {\bf D9} (1974) 1444; {\it Erratum} {\bf 10}
(1996) 1669.

\bibitem{anomdg}
R. D. Ball, S. Forte and G. Ridolfi,
\newblock Phys. Lett. {\bf B378} (1996) 255.

\bibitem{GRV}
M. Gl\"{u}ck, E. Reya and W. Vogelsang,
\newblock Nucl. Phys. {\bf B351} (1991) 579.

\bibitem{unpolCR}
R. Blankenbecler and S.J. Brodsky,
\newblock Phys. Rev. {\bf D10} (1974) 2973;\\
J. Gunion,
\newblock Phys. Rev. {\bf D10} (1974) 242;\\
G. Farrar, 
\newblock Nucl. Phys. {\bf B77} (1974) 429.

\bibitem{GSt}
T. Gehrmann and W. J. Stirling,
\newblock Phys. Rev. {\bf D53} (1996) 6100.

\end{thebibliography}
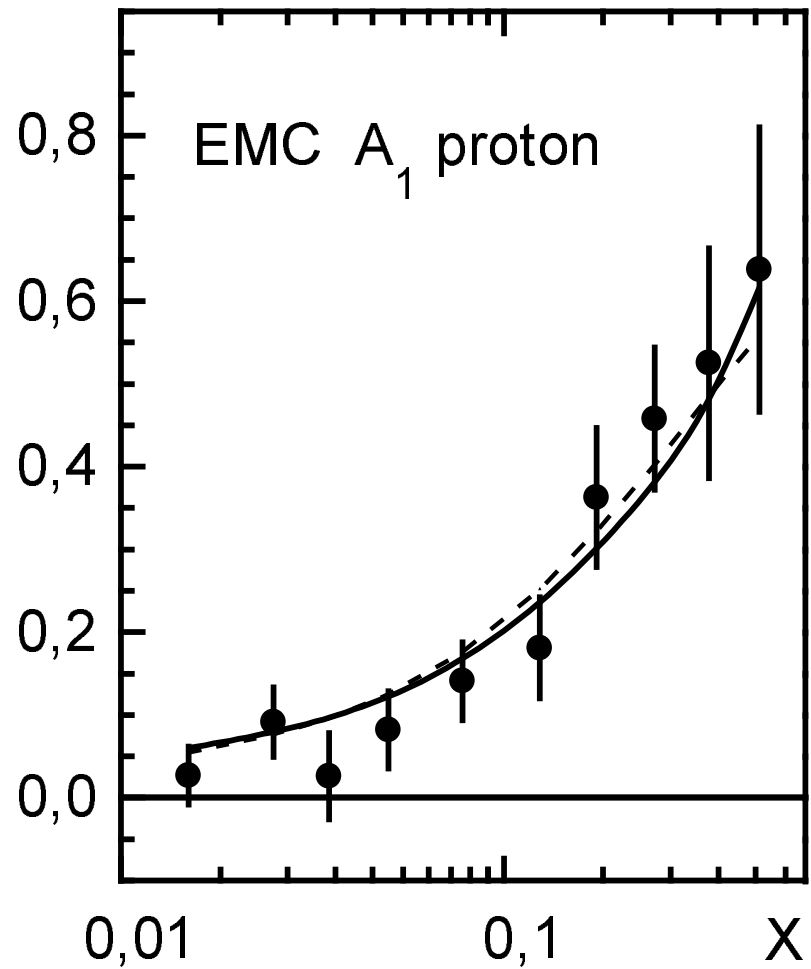
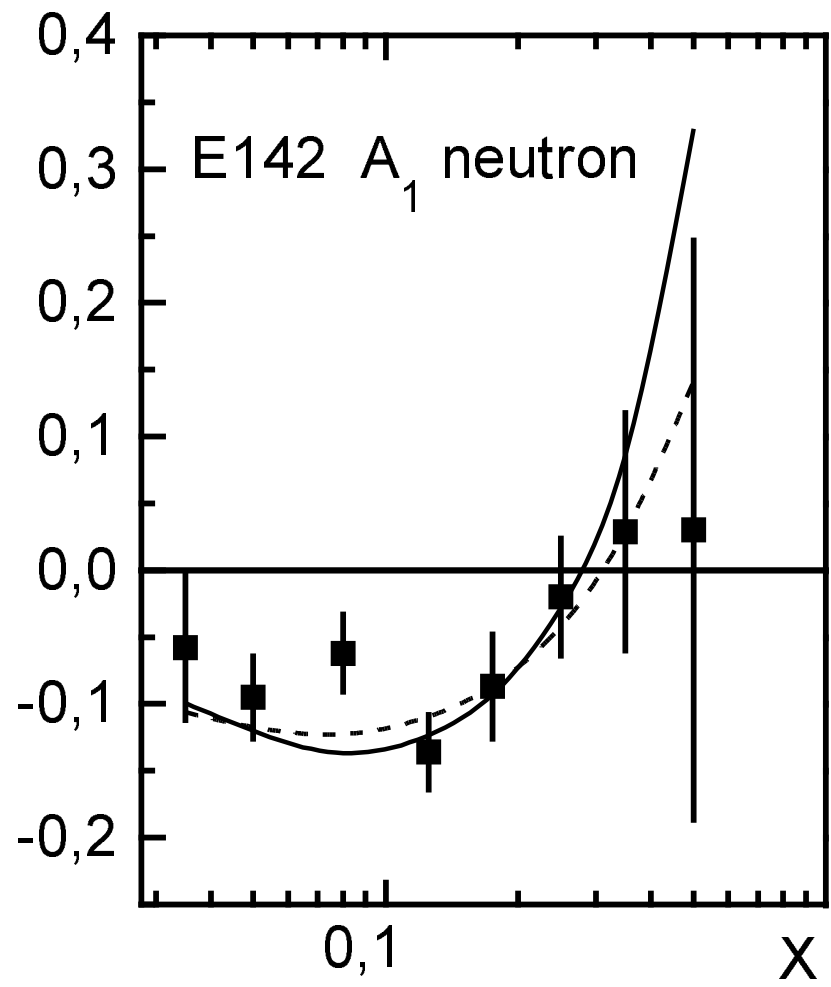
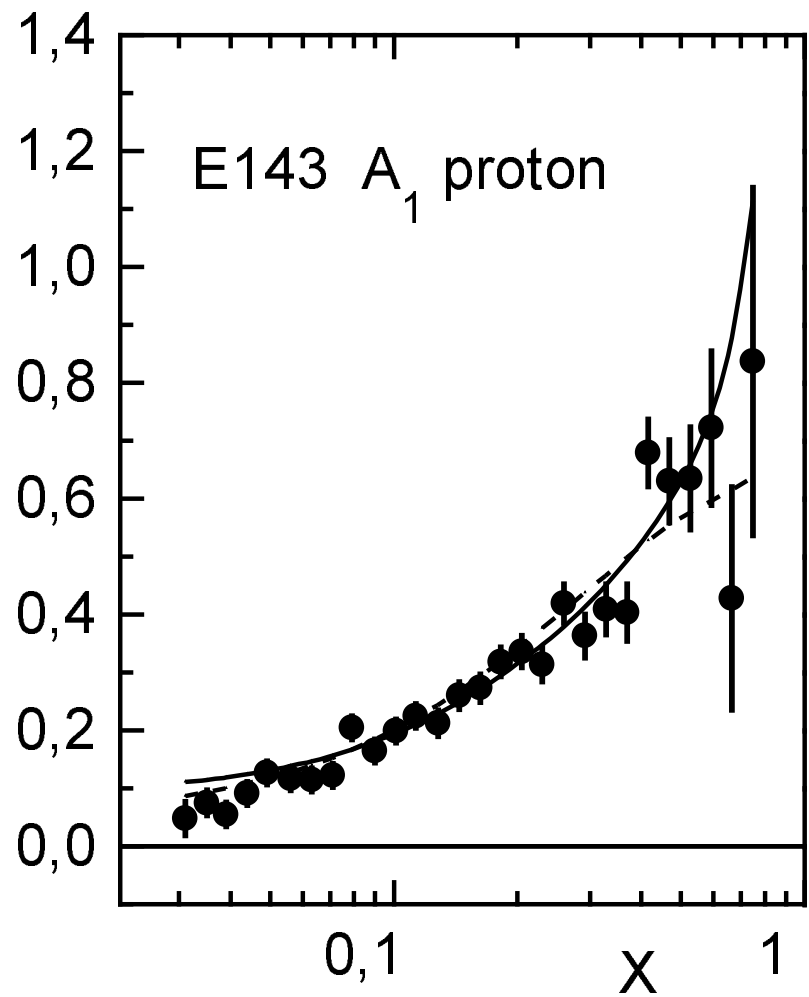
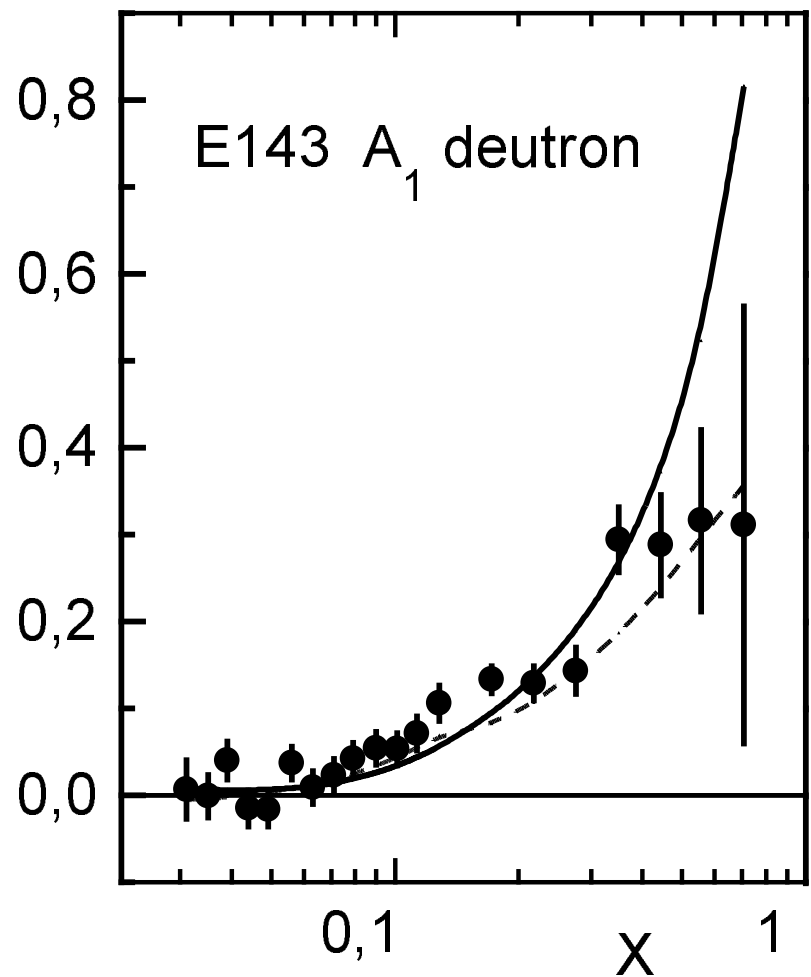
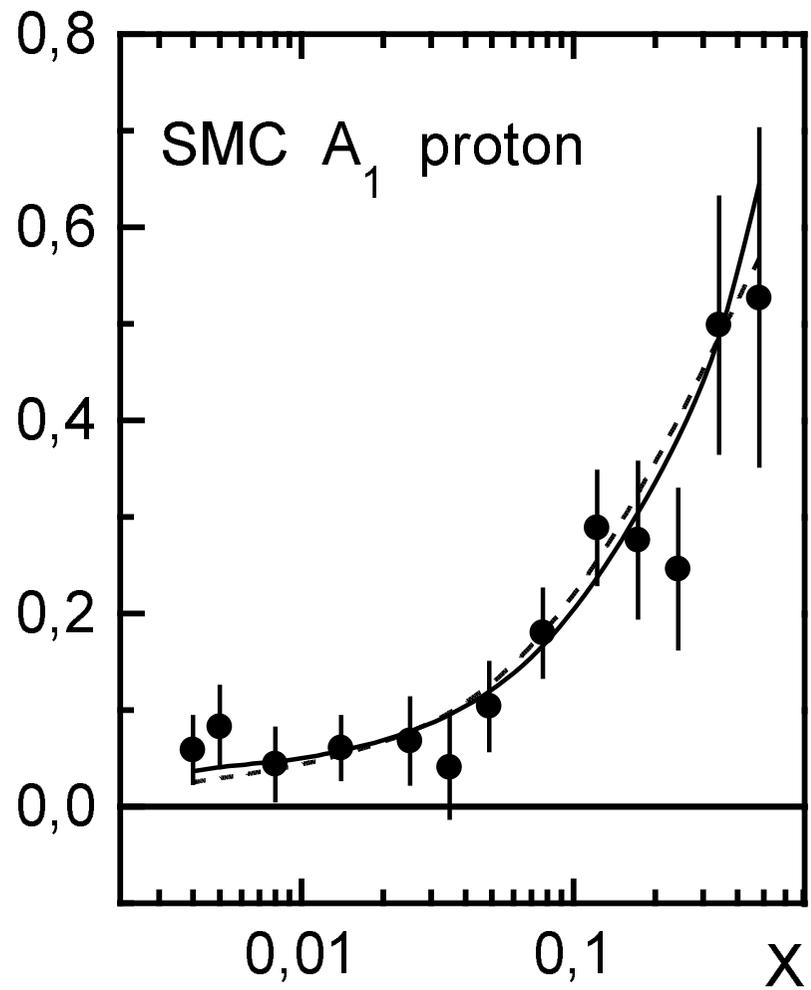
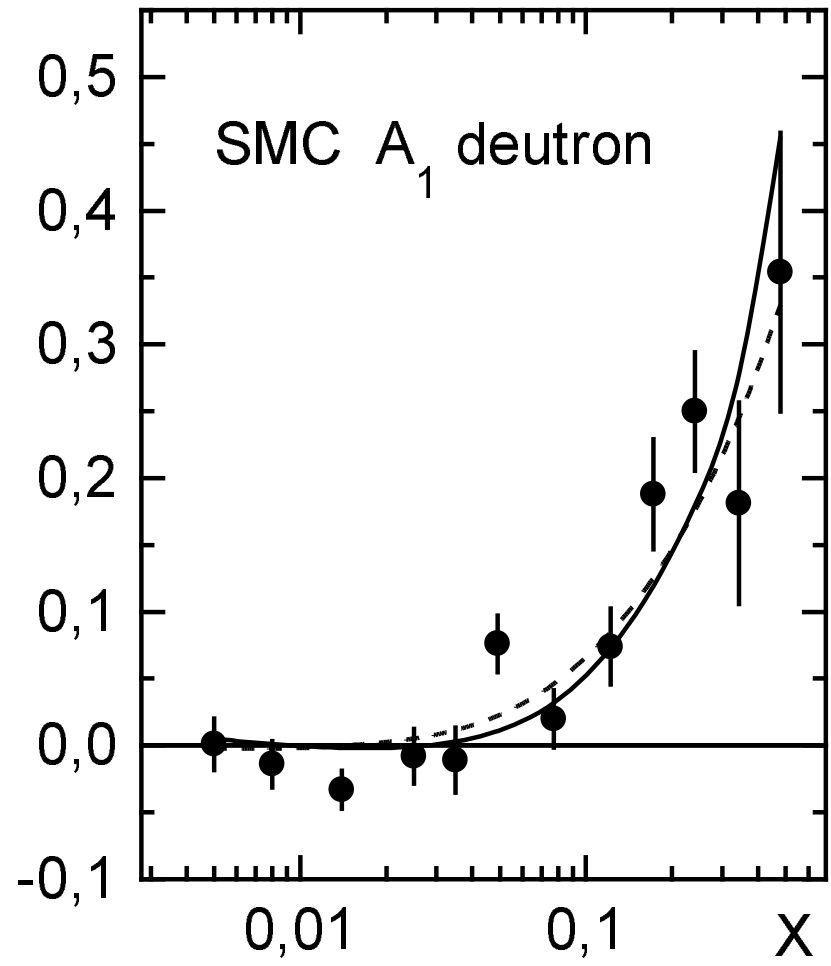
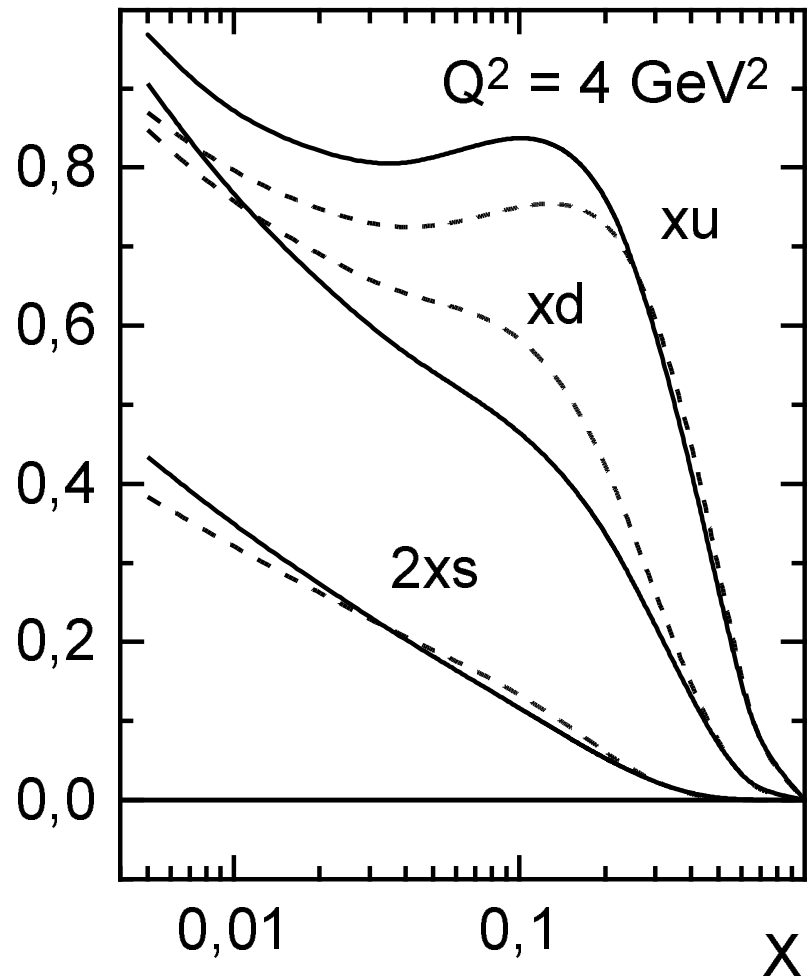
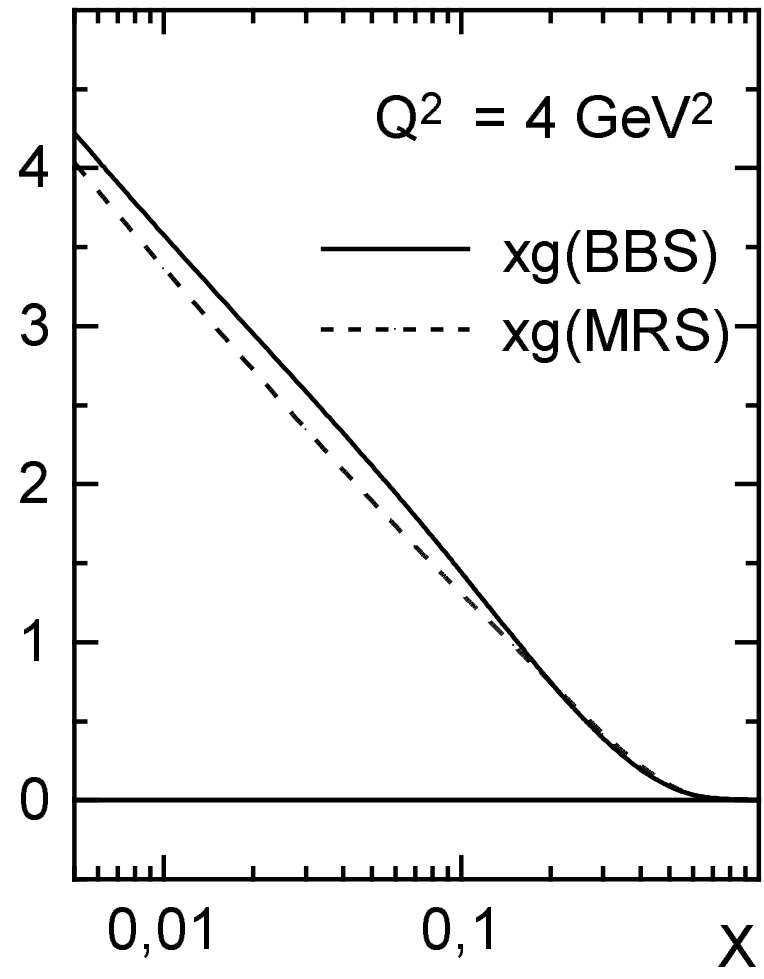
\end{document}